# Guidelines to Develop Trustworthy Conversational Agents for Children

Escobar-Planas, Marina[1,2 \[0000-0002-4513-020X\]], Gómez Emilia[2 \[0000-0003-4983-3989\]] and Martínez-Hinarejos, Carlos-D[1 \[0000-0002-6139-2891\]]

[1] Universitat Politècnica de València, Camino de Vera, s/n, 46022, València Spain
[2] European Commission, Joint Research Centre, Seville, Spain
marescplajob@gmail.com

**Abstract.** Conversational agents (CAs) embodied in speakers or chatbots are becoming very popular in some countries, and despite their adult-centred design, they have become part of children's lives, generating a need for children-centric trustworthy systems. This paper presents a literature review to identify the main opportunities, challenges and risks brought by CAs when used by children. We then consider relevant ethical guidelines for AI and adapt them to this particular system and population, using a Delphi methodology with a set of experts from different disciplines. From this analysis, we propose specific guidelines to help CAs developers improve their design towards trustworthiness and children.

## 1 Introduction

### 1.1 Motivation

Dialogue Systems, Virtual Assistants, Chatbots ... Conversational agents (CAs) have many different names, but they all refer to a computer program that supports conversational interactions with humans (McTear, 2020). Nowadays, CAs have become very popular, and recent developments allow people to interact with small computers placed in handy gadgets through voice: Google assistant can guide you in a car, Siri can send your messages in a smartphone, or Alexa can play music on a smart speaker. CAs are gaining popularity as a greater number of people are using them in their daily life.

Traditional CAs are composed of five different modules as illustrated in Figure 1: **ASR**, an automatic speech recognition engine that transforms audio speech inputs into text; **NLU**, a natural language understanding system that semantically interprets an input text; **DM**, a dialogue manager that manages the CA actions at the communication level and at the action level; **NLG**, a natural language generator that translates the computer intent into text; **TTS**, a text to speech system that transforms text into audio output. Nowadays many systems use machine learning techniques to fulfil the task of one or several modules.

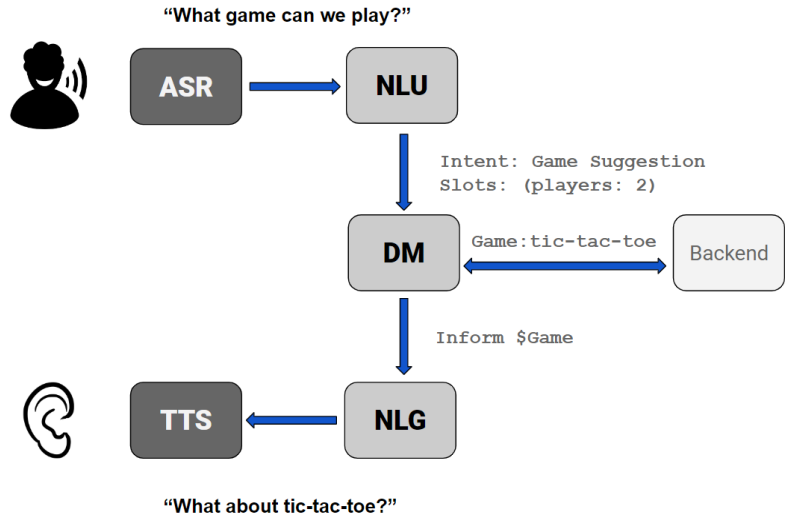

**Fig. 1. Main modules of CAs.**

Despite the adult centric design of CAs, accessibility and popularity among children should be carefully considered, since even the younger ones can interact with CAs through voice. For instance, Sciuto found out that children make a more extensive use of these devices and explore further capabilities than adults (Sciuto, 2018). This explains the huge impact CAs can have on the little ones. As an example, in a CNN article, Kelly quotes: *"The first four words my toddler understood were 'mom', 'dad', 'cat' and 'Alexa'"* (Kelly, 2018). There is then a need to research the impact of CAs on children, being a vulnerable population widely exposed to these technologies. In addition, we need some ethical guidelines for the development of CAs that can be trusted by children.

## 1.2    Goals and structure of the paper

The goal of this paper is to provide some practical ethical guidance to CAs developers, considering children as a target audience. These guidelines are intended to enhance the opportunities of CAs while minimising the risks they may bring to this vulnerable population. Our study has then two main goals: (1) Identify opportunities vs risks of CAs for children and the main ethical considerations documented in the research literature; and (2) Adapt existing ethical guidelines for Artificial Intelligence (AI) to these particular systems (CAs) and target population (children).

These two goals are addressed as follows. Section 2 presents a literature review on the opportunities, risks and challenges of developing CAs for children, and an analysis of relevant ethical guidelines in the context of AI and children's rights. Section 3 presents our approach for adapting the two considered ethical guidelines to our particular context. Section 4 presents the obtained results, further discussed in Section 5. Section 6 summarises our main conclusions and steps for future research.

## 2  Literature Review

### 2.1  Opportunities, risks and learned lessons

The research literature has identified **opportunities** brought by CAs to children, summarised as follows:

- **Improvement of accessibility**: CAs can facilitate the interaction with computers to children too young to write, with dyslexia or physical disabilities (Catania et al., 2021; Pradhan et al., 2018).
- **Engagement of learning**: CAs can support information search (Landoni et al., 2020) language learning (Nasihati et al., 2018), or school material learning (Law et al., 2020; Xu and Warschauer, 2020).
- **Promotion of social behaviour**: CAs that requires the user to use persuasive strategies in games with them (Fraser et al., 2018), or help autistic children with their social skills (Zhang et al., 2020).
- **Support of health at homes**: CAs have been used to help recording treatments and track certain diseases (Sezgin et al., 2020).

Recent studies have also identified some **risks** brought by CAs and challenges that need to be considered, mainly bias due to adult-centric design of CAs, children's overtrust and potential unexpected impact. In fact, different issues have been identified depending on different modules (Table 1).

Table 1. Problems of CAs traditional modules with children.

| Module | Children-specific characteristics |
|--------|-----------------------------------|
| ASR | Speech acoustic characteristics, e.g. high pitch range, particular prosody. |
| NLU | Expressions, vocabulary and grammar. |
| DM | Information needs, protection, allowed functionalities according to age. |
| NLG | Need of simpler words or explanations according to age. |

We have identified the following main **suggestions** to overcome the mentioned risks:

- **Communication abilities**. Children's speech and understanding should be considered in the interaction to improve children's inclusivity. Even with the current technology it is difficult to overcome these biases, however, some researchers try to improve CA's performance for children. Lavechin developed a speech identifier for babies (Lavechin et al., 2020), and other researchers identified good strategies to follow when a system does not understand a child (Cheng et al., 2018).
- **Dialogue management**. User's age should influence certain decisions of the system. Verbal and not verbal responses should also be appropriate.
- **Transparency**. Another relevant risk of CAs is to generate overtrust in children. Children tend to perceive CAs as friends, so CAs might influence them, e.g. in terms of data disclosure (Druga et al., 2020; Kahn et al., 2012). Straten taught us that transparent information helps to fight overtrust (Straten et al., 2020).
- **Continuous evaluation**. CAs are new in our lives, and the impact of their use on our society and children are still to be discovered. An example of an unexpected problem is parents asking Amazon to change the wake-up word "Alexa" in their CA product because their daughters, named Alexa, were suffering bullying at school due to this name coincidence (Johns, T., 2021). In consequence, evaluation and oversight of these devices have become highly relevant for the early detection of potential risks, and to implement the needed intervention practices.

## 2.2    Ethical guidelines for trustworthy AI and children

In recent years, several organisations have paid special attention to the ethical development of AI systems. Their aim is to generate awareness on AI systems, contribute to their understanding and evaluation, and study how to minimise the risks they can bring, while maximising their benefits. In this study, we focus on two main initiatives: the Ethical Guidelines for Trustworthy AI and UNICEF policy guidance on AI for children.

The High Level Expert Group (HLEG) of the European Commission developed the **Ethical Guidelines for Trustworthy AI** (AI HLEG, 2020), motivated by the need to protect people's fundamental rights in different contexts where AI systems are used. These ethical guidelines include seven requirements and are complemented by an assessment list for trustworthy AI (ALTAI), designed as a practical tool to help organisations self-assess the trustworthiness of their AI systems. ALTAI is a list of sixty-nine self-evaluation questions, grouped in the mentioned seven requirements as follows:

1. *Human agency and oversight* (11 questions). AI systems should respect human autonomy and decision-making, and should be supervised by humans.
2. *Technical robustness and safety* (21 questions). AI systems should be accurate, reliable and safe, having a preventative approach to risks.

3. ***Privacy and data governance*** (6 questions). AI systems should protect our privacy and have legitimate access to our data.
4. ***Transparency*** (5 questions). AI systems should have clear documentation and inform users about its decisions, capabilities and limitations.
5. ***Diversity, non-discrimination and fairness*** (10 questions). AI systems should ensure inclusion through all the AI system's life cycle.
6. ***Societal and environmental well-being*** (8 questions). AI systems should benefit the world and society.
7. ***Accountability*** (8 questions). AI systems should have mechanisms to ensure responsibility for development, deployment and use of AI systems.

The **UNICEF policy guidance on AI for children** is a guide intended to help policy makers and businesses by raising awareness of children's rights in the context of AI systems. It is proposed to complement existing work, guided by nine requirements that are presented in Table 2 (Digdum et al., 2021).

We performed a qualitative mapping between HLEG ALTAI and UNICEF AI for children requirements to understand the suitability of ALTAI having in account UNICEF consideration for AI and children's rights. From the detailed definitions, we related the 9 UNICEF requirements to the 7 requirements of HLEG. For instance, UNICEF requirement 9 refers to oversight, digital divide, and ethical development of AI from governments, that we connect to HLEG requirements on oversight, agency, fairness, and societal well/being. More details about the procedure to obtain this matrix is shared in https://github.com/mescpla/CAs4Children-ETHICOMP22.git.

We observe that ALTAI has a strong focus on the development and evaluation of specific AI devices. However, some UNICEF AI for children requirements have a strong focus on policies. Nevertheless, most requirements from UNICEF AI for children are connected to at least one major and some additional requirements of HLEG ALTAI, except for the educational aspect of requirement (8) which focuses on policies and is missing in ALTAI, which only refers to work and skills in HLEG requirement 6. In addition, all requirements present in HLEG ALTAI are covered by UNICEF AI for children guidance.

From this analysis we consider in the rest of our study the ALTAI list as a starting point, with a focus on CAs, complemented by the UNICEF guidelines on AI for children, as a complementary framework connected to children's rights, e.g. incorporating educational aspects.

**Table 2** Mapping between HLEG ALTAI and UNICEF AI for children requirements. (*), (**) and (***) indicate low, mid or high correspondence between related requirements.

| UNICEF AI for children | HLEG ALTAI | | | | | | |
|---|---|---|---|---|---|---|---|
| | Human Agency and Oversight | Technical Robustness and Safety | Privacy and Data Gov. | Transp. | Diversity, Non-discrim. and Fairness | Societal and Environm. Well-being | Account. |
| **Support children's development and well-being** *Let AI help me develop to my full potential* | | ** | ** | | ** | *** | |
| **Ensure inclusion of and for children** *Include me and those around me* | | | | | *** | | |
| **Prioritize fairness and non-discrimination for children** *AI must be for all children* | | | | | *** | | |
| **Protect children's data and privacy** *Ensure my privacy in an AI world* | ** | | *** | | | | |
| **Ensure safety for children** *I need to be safe in the AI world* | ** | *** | | | | ** | |
| **Provide transparency, explainability and accountability for children** *I need to know how AI impacts me. You need to be accountable for that* | *** | ** | | *** | ** | | *** |
| **Empower governments and businesses with knowledge of AI and children's rights** *You must know what my my rights are and uphold them* | | | | * | | *** | |
| **Prepare children for present and future developments in AI** *If I am well prepared now, I can contribute to responsible AI for the future* | * | | | * | | * | |
| **Create an enabling environment** *Make it possible for all to contribute to child centered AI* | ** | | | | ** | ** | |

## 3    Proposed methodology

From the previous literature, we propose a methodology to adapt existing ethical guidelines to the use of CAs and children and incorporate identified considerations. For that purpose, we identified prioritisation and action points from ALTAI by performing a risk level analysis for every ALTAI item (question), following the metric below:

$$\textbf{\textit{Risk}} = Likelihood \text{ x } Impact \qquad (1)$$

In order to obtain these measures (*Likelihood* and *Impact*), we have followed the Delphi method (Linstone and Turoff, 1975), conducting a survey among four experts. Then, once the experts concluded the *Likelihood* and *Impact* of every ALTAI item, we assessed the risk by the matrix approach (Kovačevićet al., 2019), identifying critical points.

### 3.1    Delphi method

To follow the Delphi method, we designed a questionnaire that asks experts to rate, for each ALTAI question, how relevant it is for children vs general population and for CAs vs AI systems in general. We use two main criteria: the likelihood or frequency of application (i.e. if the question would apply to all situations or only to certain) and the impact or relevance, using a 3 point likert scale for simplicity, given that the questionnaire has 69 questions, with 4 ratings per question (Table 3). In addition, we provided some space for experts to comment on their specificities and reasoning behind the rating when needed.

**Table 3    Example of the questionnaire**

| ALTAI   question | Dose it apply to children? (*Likelihood*) | Is it relevant for children? (*Impact*) | Does it apply to CAs? (*Likelihood*) | Is it relevant for CAs? (*Impact*) | Notes |
|---|---|---|---|---|---|
| *"Do you communicate to users that they are interacting with an AI system instead of a human?"* | Always Sometimes Never | High Medium Low | Always Sometimes Never | High Medium Low | |

We explained the questionnaire to experts from different disciplines and areas (AI ethics, CAs, education) who independently filled it using their own criteria. Individual answers were then analysed in order to identify disagreements (the mean of similar answers was 74%) and critical points. An expert meeting was later organised to discuss the identified critical points and disagreements and arrive at a common criteria and consensus. After the meeting, experts had the chance to review and refine their individual responses and submit their final version (84% similar answers).

## 3.2 Risk assessment

For each of the ALTAI items (i.e. questions) we computed partial risk levels for children and CAs, as follows:

**Risk value for ALTAI questions**. For each item of ALTAI, we performed the arithmetic mean to combine the four reported *Likelihood* and *Impact* measures from the experts related to children and CAs in a separate way. Later on, we built the *Child Risk* and the *CA Risk* (called partial risks from now on) by using Formula (1) with their respective *Likelihood* and *Impact* means. In addition, a risk assessment matrix was built to measure the level of risk of our results (Fig.2.a).

**Risk value for HLEG ALTAI requirements**. We computed the arithmetic mean for every question inside a given requirement (e.g. *Human Agency and Oversight* requirement is composed of eleven questions), to calculate the *Likelihood* and *Impact* of the requirement. We did it separately to calculate the partial risks for children and CAs. Later on, the *Child Risk* and the *CA Risk* of a particular requirement was calculated following Formula (1).

From individual partial risks (*Child Risk* and *CA Risk*) per question and requirement, we calculate the *Total Risk* of every question and requirement by the following formula:

$$\textbf{\textit{Total Risk}} = \textit{Child Risk} \text{ x } \textit{CA Risk} \qquad (2)$$

Finally, we calculate a risk assessment matrix for the *Total Risk* (Figure 2.b) in order to understand the severity of the risk levels. Detailed results can be found in https://github.com/mescpla/CAs4Children-ETHICOMP22.git.

## 3.3 Risk assessment

In order to complement the quantitative assessments, we combined all notes provided by the experts in the questionnaire and the critical points discussed during the Delphi meeting. We carried out a thematic analysis (Braun & Clarke, 2006). First, we compiled all the annotated comments, identifying initial ideas. Secondly, we grouped the ideas in possible themes that we discussed and refined until our final version. Finally, we selected the examples that we would use in the report.

# 4    Results

## 4.1    Ordered assessment list

Table 4 shows the ratings for likelihood, impact and risk for children and CAs dimensions as well as total values. We first observe that, in general, the obtained *Impact* values are higher for children than for CAs. However, the *Likelihood* is much larger for CAs than for children. This explains why the *CA Risk* is higher than *Children Risk* for every category but for: *Human oversight*, *Accuracy*, *Explainability*, and *Environmental well-being* (Table 4).

**Table 4** Values for Likelihood, Impact for calculating the Child Risk, the CA Risk and the Total Risk, per each HLEG ALTAI requirement and sub-requirement. Colours indicate the risk level using the colour scheme presented in Fig.2.

| | Children | | | | | CAs | | | | | Total Risk |
| --- | --- | --- | --- | --- | --- | --- | --- | --- | --- | --- | --- |
| | Likelihood | | Impact | | Child Risk | Likelihood | | Impact | | CA Risk | |
| | Mean | SD | Mean | SD | | Mean | SD | Mean | SD | | |
| **HUMAN AGENCY AND OVERSIGHT** | **2,00** | **0,45** | **2,61** | **0,05** | **5,21** | **2,33** | **0,16** | **2,62** | **0,68** | **5,87** | **30,59** |
| Human Agency | 2,15 | 0,50 | 3,00 | 0,00 | 6,45 | 2,87 | 0,23 | 2,93 | 0,94 | 8,41 | 54,24 |
| Human Oversight | 1,80 | 0,40 | 2,13 | 0,12 | 3,84 | 1,80 | 0,00 | 2,00 | 0,40 | 3,60 | 13,82 |
| **TECHNICAL ROBUSTNESS AND SAFETY** | **1,54** | **0,46** | **2,19** | **0,16** | **3,38** | **2,03** | **0,33** | **2,02** | **0,45** | **4,10** | **13,85** |
| Resilience to Attack and Security | 1,38 | 0,58 | 2,06 | 0,19 | 2,83 | 2,11 | 0,19 | 2,06 | 0,48 | 4,34 | 12,27 |
| General Safety | 1,40 | 0,20 | 2,13 | 0,23 | 2,99 | 1,87 | 0,23 | 2,00 | 0,41 | 3,73 | 11,15 |
| Accuracy | 1,90 | 0,36 | 2,40 | 0,23 | 4,56 | 2,13 | 0,35 | 1,93 | 0,46 | 4,12 | 18,81 |
| Reliability, Fall-back plans and Reproducibility | 1,53 | 0,68 | 2,20 | 0,00 | 3,37 | 2,00 | 0,58 | 2,07 | 0,46 | 4,13 | 13,94 |
| **PRIVACY AND DATA GOVERNANCE** | **1,83** | **0,76** | **2,72** | **0,38** | **4,99** | **2,67** | **0,48** | **2,61** | **0,78** | **6,96** | **34,75** |
| Privacy | 1,88 | 0,89 | 2,67 | 0,58 | 5,00 | 2,83 | 0,29 | 2,67 | 0,84 | 7,56 | 37,78 |
| Data Governance | 1,81 | 0,69 | 2,75 | 0,29 | 4,98 | 2,58 | 0,58 | 2,58 | 0,74 | 6,67 | 33,26 |
| **TRANSPARENCY** | **1,90** | **0,40** | **2,40** | **0,23** | **4,56** | **2,27** | **0,35** | **2,27** | **0,59** | **5,14** | **23,43** |
| Traceability | 1,00 | 0,00 | 2,00 | 0,00 | 2,00 | 2,67 | 0,58 | 2,00 | 0,59 | 5,33 | 10,67 |
| Explainability | 2,00 | 0,50 | 2,50 | 0,58 | 5,00 | 2,00 | 0,00 | 2,17 | 0,48 | 4,33 | 21,67 |
| Communication | 2,25 | 0,50 | 2,50 | 0,00 | 5,63 | 2,33 | 0,58 | 2,50 | 0,69 | 5,83 | 32,81 |
| **DIVERSITY, NON-DISCRIMINATION AND FAIRNESS** | **1,75** | **0,44** | **2,37** | **0,40** | **4,14** | **2,63** | **0,46** | **2,17** | **0,64** | **5,71** | **23,63** |
| Avoidance of Unfair Bias | 2,00 | 0,53 | 2,53 | 0,46 | 5,07 | 2,80 | 0,35 | 2,33 | 0,73 | 6,53 | 33,10 |
| Accessibility and Universal Design | 1,31 | 0,33 | 2,17 | 0,29 | 2,84 | 2,50 | 0,58 | 1,92 | 0,53 | 4,79 | 13,63 |
| Stakeholder Participation | 2,25 | 0,50 | 2,33 | 0,58 | 5,25 | 2,33 | 0,58 | 2,33 | 0,60 | 5,83 | 28,58 |
| **SOCIETAL AND ENVIRONMENTAL WELL-BEING** | **1,27** | **0,53** | **2,04** | **0,89** | **2,59** | **1,50** | **0,94** | **2,00** | **0,33** | **3,00** | **7,78** |
| Environmental Well-being | 1,00 | 0,00 | 3,00 | 0,00 | 3,00 | 1,67 | 1,53 | 1,33 | 0,25 | 2,22 | 6,67 |
| Impact on Work and Skills | 1,43 | 0,84 | 1,47 | 1,42 | 2,10 | 1,33 | 0,69 | 2,20 | 0,33 | 2,93 | 6,17 |
| Impact on Society at large or Democracy | 1,00 | 0,00 | 3,00 | 0,00 | 3,00 | 1,00 | 1,00 | 2,33 | 0,52 | 4,67 | 14,00 |
| **ACCOUNTABILITY** | **1,22** | **0,34** | **2,54** | **0,51** | **3,10** | **2,42** | **0,83** | **2,04** | **0,55** | **4,93** | **15,28** |
| Auditability | 1,00 | 0,00 | 2,17 | 0,29 | 2,17 | 3,00 | 0,00 | 2,00 | 0,67 | 6,00 | 13,00 |
| Risk Management | 1,29 | 0,45 | 2,67 | 0,58 | 3,44 | 2,22 | 1,10 | 2,06 | 0,51 | 4,57 | 15,73 |

Considering the partial risk assessment matrix (Fig.2.a), our experts have identified *Human agency and oversight*, *Privacy and data governance* and *Transparency* as the main critical points for children, while *Privacy and data governance*, *Human agency and oversight*, and *Diversity, non-discrimination and fairness* are the main critical requirements for CAs. Regarding the *Total Risk*, and considering the combined risk assessment matrix (Fig.2.b), we identify *Privacy and data governance* and *Human agency and oversight* (with a critical point on *Human agency*) as the main critical requirements to be considered when developing CAs for children. The only requirement which values are over the matrix diagonal is *Societal and environmental well-being,* with the lowest partial and combined risk levels scores.

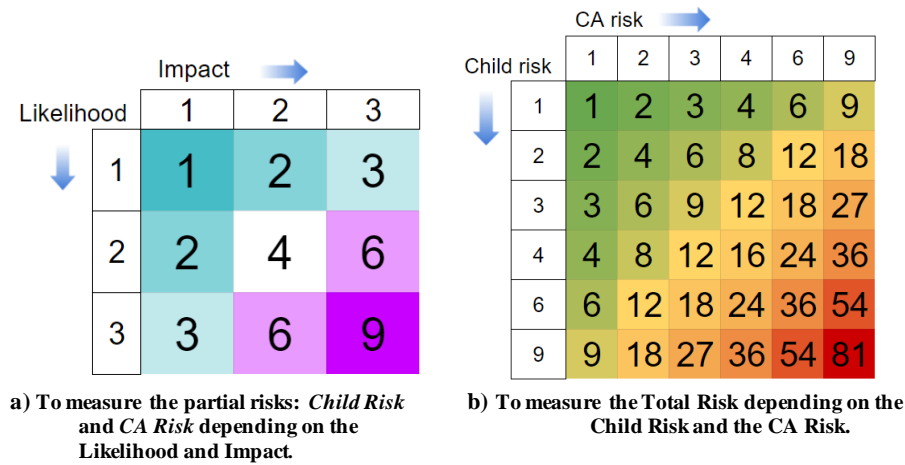

a) To measure the partial risks: *Child Risk and CA Risk* depending on the Likelihood and Impact.

b) To measure the Total Risk depending on the Child Risk and the CA Risk.

**Figure 2** Risk assessment matrices.

## 4.2 Thematic Analysis

During the Delphi meeting, our experts (R1, R2, R3 and R4) discussed some relevant topics, and other critical considerations were pointed out as questionnaire notes (Table 3). As mentioned in the methodology (Section 3), we performed a thematic analysis and identified critical considerations for the ethical design of child-centric CAs. These considerations are visually presented in Figure 3 and Table 5, and summarised as follows:

**Involve children stakeholders.** Many of the experts commented on the relevance of involving children stakeholders (children, teachers, parents …) in the design, use, and test of the system. R1 wrote: "*Include children, tutors and teachers as stakeholders*", R2 mentioned: "*Multiple stakeholders need to be involved*". Furthermore, it was mentioned the need to *"teach stakeholders"* (R2), so they can help to oversee the system. In addition the experts agreed that -as children must not work- their collaboration in the design, use, and test of the system needs to be done in a meaningful and entertaining way. As R4 said: *"We need to involve children in the design, but in a meaningful way as this participation should be far from job conditions".*

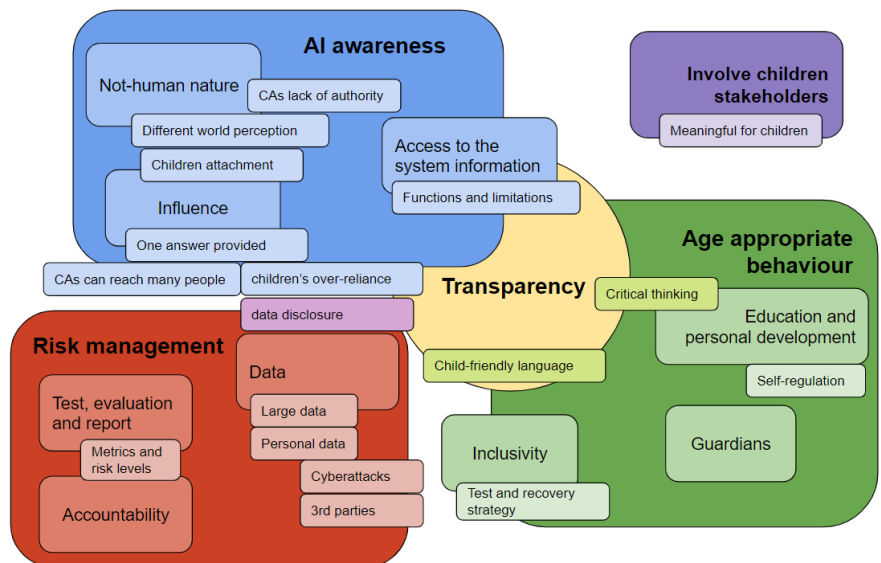

**Figure 3 Annotation scheme of experts' comments.**

**AI awareness.** The experts expressed many considerations regarding the fact that people -particularly children- should be aware of what a CA is, how it works and its limitations. In particular:

**Not-human nature**. The experts remarked that CAs communicate in a natural way, which can lead to confusion about their nature (*"Naturalness of CAs might create confusion"* (R2)). Therefore, there is a risk of developing attachment to the device, especially for children developing their social abilities and with a particular way to understand the world. R4 wrote "*Careful with human attachment as children are developing their cognitive and emotional abilities*", R1 also shared: "*Children can think that something that is not alive has alive characteristics such as feelings*". In addition, the social role of the device should be regarded, as mentioned by R4: "*Children might understand what the system is and that has no authority*". The experts also discussed the positive points of being consistent with the information provided by the CA "*If we want a child to understand that a CA has no feelings, maybe it is better to avoid sentences such as 'I am happy'*".

**Influence**. Another critical consideration is the influence a system might have on the user. Firstly, R3 remarked *"CAs can provide information critical to make decisions. People don't usually double check information"*. Additionally, R4 shared concerns about children's critical thinking "*Children need to learn to be critical, to look further from the provided information, to develop creativity*". Secondly, regarding CA's influence, the experts expressed their concerns about children's vulnerability: "*Check over-reliance. Consider children's vulnerability*" (R2); "*Special attention to kids over-reliance*" (R3). Children are highly impressionable -even more provided that we consider the previous point- and their judgement and awareness, on things such as data disclosure, are still in development. Following this, R4 pointed out: "*Kids don't yet have good judgement, careful with influence and over-reliance*". Last but not least, R2 highlighted the relevance of checking a CA's influence, as their impact is magnified by the number of people able to use the device: "*CAs can influence a large number of people, so the social impact of CAs should be considered*".

**Access to the system information.** From a general point of view, all our experts advised developers to be transparent about the nature, functioning and limitations of the system. This information promotes AI awareness, and can minimise some of the above mentioned risks. R3 also commented: "*Understanding CAs nature and interactions is important to avoid frustration*". In addition, during the experts' meeting, they discussed that this information should be always accessible to the user, and provided in a proactive manner by the system in accordance with the duration and risks of the interaction. R1 mentioned: "*Regarding explainability it should also fit the purpose of the CA. It doesn't make sense to explain all the limitations of a system that will interact with you for a few seconds (e.g. to ask you on the phone what department you want to contact).*" R3 added that this would be necessary in bank transactions or high risk interactions.

**Transparency** (e.g. informing on how an AI system works, and what are its limitations) is a transversal topic mentioned in different requirements. It is considered as a tool to fight different risks brought by AI, by raising AI awareness and enhancing critical thinking. The access to information about the system (i.e. working principles and limitations), is then essential. Two clear examples of the use of information to fight data disclosure are the "*relevance to inform about recordings*" (R3); and being "*important to know if the system learns about them or not*" (R4). Moreover, all the experts highlighted the need to communicate with children in a child-friendly language ("*Questionnaires or explanations must be adapted to children*" (R1)).

**Risk Management.** Regarding the novelty of CAs in people's lives, it is paramount to detect potential risks and ensure the accountability of the developed systems. It is also essential to understand which personal data a CA is using and storing, as related to the privacy requirement. Some identified critical aspects of risk management:

**Test, evaluation and reporting.** Many CAs tend to have a low accuracy with small children and other vulnerable groups. In addition, risks for children require continuous vigilance. Defining metrics and levels to handle risks can help with this problem; but, in addition, the experts suggested some strategies to minimise these risks. On the one hand, developers can evaluate the system in a controlled environment, involving children and other minorities (*"Relevance of testing and detect if a CA is having problems with children"* (R3); *"Test the system for different children considering vulnerable groups"* (R4)). On the other hand, developers can provide users - including children- with mechanisms to report issues that appear after extensive use in the real world scenarios (*"Flag issues in a child-friendly way"* (R4); *"Children should be able to report issues"* (R2)).

**Accountability.** The experts shared some concerns about the vulnerability of children, and the relevance of people taking responsibility for the system. For instance, R2 wrote: "*Children are a vulnerable population in developmental stage*", and R3 shared: "*As CAs use biometrical data, and children are a vulnerable population*, a special attention to accountability is needed". Audits may help to keep track of this problem.

**Data.** Regarding data, the experts raised some concerns. First of all, they emphasise on the recording of biometric data by CAs (*"CAs data storage contain biometrics and personal data"* (R2); *"CAs use biometric data"* (R3); *"Voice is personal data"* (R4)) and children's vulnerability (**"***Children have the right to be protected"* (R1); *"Extra protection for vulnerability, careful with children over trust"* (R4)). Therefore, not only should we make an extra effort for personal data protection, (compliance with European regulations such as *"GDPR (*General Data Protection Regulation*)"* (R4)), but we should also pay a special attention, if the data is shared or transferred to 3rd parties (*"Careful about selling data to 3rd parties"* (R4)), and protect the system from cyberattacks (*"CAs have critical data and must be protected from cyberattacks"* (R3)). Secondly, the consulted experts highlighted that some CAs use information from large data sources (e.g. CAs can search for information on the internet, or can use large datasets for training using deep learning). Among the risks of using these data sources, they mentioned the lack of control of possible risks that can later appear in the system (*"CAs with untrusted data sources may cause more damage"* (R3). For instance, the experts mentioned in the meeting that, last summer, a game displayed on a CA that took information from the web, told a child to put a coin on a connected plug. The experts recommend a better risk management for those systems.

**Age appropriate behaviour.** During the study our experts also identified the need to recognise children, and act appropriately, i.e specific considerations/behaviours when interacting with children:

**Inclusivity.** CAs bring opportunities for inclusivity of illiterate people, or people with disabilities (*"CAs help with inclusivity and should take special care on this point, special attention to disabilities"* (R3)). That is why developers should emphasise on the inclusivity of the device, setting an example to children (*"Children can internalise bias, so it is important for* *them"* (R3); *"Children inclusivity for all culture, language, age, .."* (R2)). Nonetheless, these systems face some challenges understanding little children and underrepresented groups (*"Biases linked to not available data, which is a challenge for all EU languages, dialects and children"* (R4); *"Consider limitations of CAs understanding different people"* (R1)). Consequently, our experts remarked the importance of recognizing children as users, and to keep working against this bias (*"Special attention to bias towards children"* (R1); *"Discrimination by age", (*R2)). In the meantime, a good strategy to fix conversations when the system cannot identify the user might help to mitigate this risk ("*Important to use a good recovery strategy*" (R1)).

**Guardians.** Children have a particular autonomy, as they need the supervision of their guardians. This should be taken into consideration when designing a device that can interact with a child. The experts highlighted their presence in different points. They advise to consider them in order to meet consent obligations (*"Need of tutor consent"* (R4); **"Tutors and children must give their consent"** (R1)), but also to take advantage of their presence during interactions (*"Rely on adult supervision when low confidence"* (R2); *"Children are not aware, so an adult should supervise"* (R3)). During the experts' meeting, they discussed the supervision of guardians, identifying their presence, but bearing in mind that the system should be safe enough for the child to not require constant supervision. They agreed that the system can try to use them/call them in specific moments, although the security of the system cannot be just based on the guardians' supervision.

**Educational and personal development.** All the experts missed a section on education and children development. For instance R1 commented: *"We need to consider CAs in education"*, R3 mentioned: *"Need for educational consideration"*, and R2 wrote: *"We should consider adding to ALTAI education and development questions"*, referring not just to school education, but also to personal development such as self-regulation (*"Consider children addictive behaviour"*, R3).

The experts' recommendations cover critical points for all the seven requirements from HLEG ALTAI (Table 5). Being *Societal and environmental well-being* the requirement with more critical themes pointed out by the experts, followed by *Human agency and oversight* and *Technical robustness and safety*. Experts also covered all the learned lessons from the literature review.

**Table 5 Thematic analysis mapped to the requirements of HLEG ALTAI.**

| | Agency | Robust. | Data | Transp. | Diversity | Well-being | Account |
|---|---|---|---|---|---|---|---|
| **Involve children stakeholders** | | x | | x | x | x | |
| **AI awareness** | x | x | x | x | | x | |
| Not-human nature | x | | | x | | x | |
| Influence | x | | | | | x | |
| Access to the system information | x | | x | x | | x | |
| **Transparency** | x | | | x | | x | |
| **Risk management** | | x | x | | x | | x |
| Test and report | | x | | | x | | |
| Accountability | | | | | | | x |
| Data | | x | x | | | | |
| **Age appropriate behaviour** | x | x | x | x | x | x | |
| Education and personal development | x | | | | | x | |
| Guardians | x | x | x | x | | x | |
| Inclusivity | | x | | | x | | |

## 5  Discussion and recommendations for child-centric CA developers

From the results presented in Section 4, we recommend developers to consider the ALTAI assessment list, in the **following order of priority** (Fig.3): *Privacy and data governance*, *Human agency and oversight*, *Diversity, non-discrimination and fairness*, *Transparency*, *Accountability*, *Technical robustness and safety*, and *Societal and environmental well-being*. In addition, developers should pay special attention to the considerations outlined by the experts (Fig.6): *Involve children stakeholders*, *AI awareness*, *Transparency*, *Risk management* and *Age appropriate behaviour*. These recommendations will help developers to maximise CAs opportunities for children while minimising risks, creating more accessible CAs, supporting educational activities, social behaviour and safety.

Another interesting conclusion from our work is the identification of a subsection that could enrich current ALTAI guidelines for children: an **"education and self-development"** set of questions in the *Societal and environmental well-being* requirement.

Furthermore, in our risk assessment analysis, the **main critical point detected is *Privacy and data governance*.** This point was also covered by our experts' critical considerations in the identified topics of *Risk management* and *Age appropriate behaviour* where they highlighted the presence of children's guardians. These considerations are aligned with previous studies (von Struensee, 2021). Nevertheless, while the use of data protection regulations is well established, we found little research on the application of data privacy regulations considering AI, children autonomy, and guardians. Therefore, we recommend integrating research outcomes from existing medical studies that use biometric data from children (Hopf et al., 2014).

Besides, our results bring special attention to *Human agency and oversight*, with a special focus on the **not-human nature of the system.** This was also reflected on the thematic analysis in AI awareness using *Transparency* as a tool. These recommendations are in accordance with existing work (Straten et al., 2020).

We also identified some **limitations** of our study. Firstly, our study comes from an european-centric perspective, e.g. focus on Ethical guidelines for trustworthy AI and involving an EU expert group. Therefore, these results cannot then be generalised to different cultures such as Asian or African. Henceforth, we suggest complementary research in larger and more diverse groups with different cultural backgrounds.

Secondly, regarding our metrics, *Children Risk* was generally lower than *CAs Risk*, mainly because of a high rated CAs *Likelihood*. We recognize that our metrics put at the same level children and CAs considerations, when it might be more adequate to highlight children's considerations, e.g. through weights. We encourage complementary studies with alternative metrics.

Finally, HLEG ALTAI *Societal and environmental well-being* requirement had the lowest risk level - surprisingly low considering it is also a fundamental requirement on other studies on AI and children's rights (Charisi, et al., 2022). This might be due to the inclusion of a work impact section (with a low *Likelihood* and *Impact* on children), and the lack of an *Education and self-development* section. This new - and needed- section would change our results of the risk assessment. We encourage further research to build new items for ALTAI on this topic, and suggest the use of LifeComp competences for its development (Sala, et al., 2020): personal (self-regulation, flexibility, well-being), social (empathy, communication, collaboration), and learning to learn (growth mindset, critical thinking, managing learning).

# 6     Conclusions

We have performed a literature review on conversational agents (CAs), identifying opportunities, challenges and risks for their use with children. In addition, we have consulted a group of experts to measure the risk of all the items of the assessment list for trustworthy Artificial Intelligence (ALTAI) with a focus on children and CAs. With our results, we adapt ALTAI for this specific use, defining priorities on the requirements and adding additional considerations. We hope this research can help CA developers to build trustworthy child-centric systems that can respect fundamental and children rights, ensuring a future where children can also take the most from CAs. We may have safer and better-informed citizens with critical thinking in the future.

# 7    Acknowledgements

We thank our experts for their patience and dedication to our experiment, particularly Riina Vuorikari. We thank Isabelle Hupont-Torres and Marta Rivera for their guidance regarding the analysis and assessment of risk. Finally, we want to thank the HUMAINT team for their constant support and useful comments to the text.